# Suppression of secondary phase formation in Fe implanted ZnO single crystals

K. Potzger,[a] Shengqiang Zhou, H. Reuther, K. Kuepper, G. Talut, M. Helm, and J. Fassbender

Institute of Ion Beam Physics and Materials Research, Forschungszentrum Dresden-Rossendorf, P.O. Box 510119, 01314 Dresden, Germany

J. D. Denlinger

Advanced Light Source, Lawrence Berkeley National Laboratory, Berkeley, California 94720

Unwanted secondary phases are one of the major problems in diluted magnetic semiconductor creation. Here, the authors show possibilities to avoid such phases in Fe implanted and postannealed ZnO(0001) single crystals. While $\alpha$-Fe nanoparticles are formed after such doping in as-polished crystals, high temperature (1273 K) annealing in $O_2$ or high vacuum before implantation suppresses these phases. Thus, the residual saturation magnetization in the preannealed ZnO single crystals is about 20 times lower than for the as-polished ones and assigned to indirect coupling between isolated Fe ions rather than to clusters. ©2007

Diluted magnetic semiconductors (DMSs) such as transition metal (TM) doped ZnO have recently attracted huge attention due to their application potential in spintronics.[1,2] Especially for rather easy available *n*-type ZnO, TM dopants such as Fe or Co but not Mn are theoretically predicted to yield ferromagnetic coupling.[2] One of the major drawbacks in preparation is the unwanted formation of magnetic secondary phases for high TM concentrations (~5%) necessary[3,4,5] mimicking a room-temperature DMS. In this letter, we show that unwanted secondary phases in ZnO single crystals implant doped with Fe can be avoided by annealing the crystals prior to implantation. Moreover, weak ferromagnetic properties are introduced that are not related to ordinary superparamagnetic nanoparticles. Thus, the following sample set has been prepared from hydrothermal, commercial epipolished ZnO(0001) substrates purchased from Crystec: (1) nonpreannealed, i.e., as polished, (2) $O_2$ preannealed in flowing $O_2$ at 1273 K for 15 min, and (3) vacuum preannealed in high vacuum (base pressure $<1\times10^{-6}$ mbar) at 1273 K for 15 min. $O_2$ annealing at high temperatures is known to reduce lattice damage in the near surface region of ZnO.[6,7] Vacuum annealing (3), was chosen due to the formation of point defects that might mediate indirect ferromagnetic coupling of the implanted ions.[8,9,10] Note that mild vacuum annealing around 873 K usually introduces both O vacancies and Zn interstitials.[11] After high temperature annealing, however, Zn interstitials are not stable, i.e., the defects are dominated by oxygen vacancies.[12] Following that paper, oxygen vacancies are not expected to mediate ferromagnetic coupling, while Zn interstitials are. Thus, our approach, in addition to the suppression of secondary phases, would give a confirmation of these different effects of various kinds of defects for the case of Fe doping. For further processing, our samples were subjected to $^{57}$Fe ion implantation. The implantation energy was 80 keV at an incident angle of 7° yielding a projected range of 38 nm and a straggling of 17 nm (TRIM). The implanted Fe fluence of $2\times10^{16}$ cm$^{-2}$ yielded a maximum atomic concentration of 5%. In order to avoid magnetic secondary phases already in the as-implanted samples, a low implantation temperature of 253 K was used.[3] Postimplantation annealing for lattice recovery was performed in high vacuum at a temperature of 823 K for 15 min. The base pressure was below $1\times10^{-6}$ mbar. The particular parameters for the postannealing have been chosen to avoid long-range diffusion and oxidation of the implanted Fe as have been observed earlier for higher annealing temperatures.[4] For a detailed analysis we applied x-ray diffraction (XRD) using a Siemens D 5005 diffractometer equipped with a Göbel mirror for enhanced brilliance, Rutherford backscattering/channeling (RBS/C), atomic force microscopy (AFM), conversion electron Mössbauer spectroscopy (CEMS) at room temperature, x-ray absorption spectroscopy (XAS) performed at beam line 8.0.1 of the Advanced Light Source in Berkeley, and superconducting quantum interference device (SQUID) magnetometry with the magnetic field applied parallel to the sample surface. RBS/C revealed no significant change of the crystallinity after preannealing. In contrast, AFM (not shown) reveals pronounced changes of the crystal surface morphology. After O$_2$ preannealing, the root mean square surface roughness ($R_q$) of the ZnO sample slightly increases from 0.23 to 0.27 nm and regularly oriented steps appear. The latter is an indication for surface recrystallization.[7] Vacuum preannealing, in contrast, yielded a surface roughness of 23 nm. While after implantation a slight increase of $R_q$ is detectable, postannealing does not change $R_q$ significantly for any of the samples. RBS/C for both the preannealed and nonpreannealed crystals [Fig. 1(a)] shows a decrease of the minimum channeling yield ($\chi_{min}$) with postannealing. The drop is largest for the nonpreannealed crystal and smallest for the vacuum preannealed sample. The lowest $\chi_{min}$ is achieved for the O$_2$ preannealed crystal. $\chi_{min}$ directly reflects the crystalline homogeneity, i.e., while an amorphous sample shows a $\chi_{min}$ of 100%, a perfect single crystalline sample exhibits 1%–2%. Diffusion of the implanted Fe due to postannealing could be ruled out by means of RBS/C random spectra [Fig. 1(b)]. Bumps originating from the implanted Fe are visible in all samples that allow us to investigate the potential diffusion of the Fe inside ZnO. Upon postannealing at 823 K these bumps do not shift, i.e., Fe is not segregating over larger distances. In order to check the potential formation of secondary phases, XRD of the implanted and postannealed samples has been performed. The presence of secondary phases has only been observed for the nonpreannealed and postannealed crystal (not shown), i.e., $\alpha$-Fe nanoparticles of 7 nm mean diameter, as calculated using the Scherrer formula.[13]

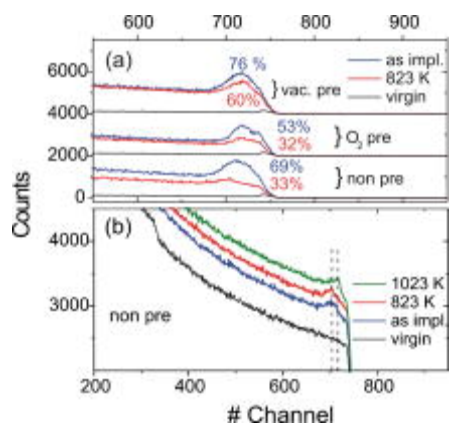

Fig. 1. (Color online) (a) Channeling RBS spectra for the as-implanted and postannealed ZnO single crystals. Preimplantation annealing is indicated. The $\chi_{min}$ value for every spectrum is given in %. (b) Exemplary random spectra for the nonpreannealed crystal for different postannealings. Fe is visible in the random spectra as a bump. All spectra have been shifted in $y$ direction for better clarity.

In order to further prove that metallic nanoparticle formation has been avoided by our preannealing, element specific spectroscopy was applied. We performed CEMS and XAS, respectively. While CEMS is more bulk sensitive, XAS recording the total electron yield is rather sensitive to the surface region. The combination of both methods thus leads to a complete picture of the charge states of the implanted Fe. Moreover, CEMS allows simultaneous detection of electronic and magnetic properties at the nucleus of the implanted Fe. The CEM spectra of the as-implanted samples (not shown) look similar exhibiting mixed $Fe^{2+}$ and $Fe^{3+}$ valence states. No magnetic sextet was detected for any of the samples. Thus, they are comparable to the ones from earlier work.[3,14] Figure 2 shows CEM spectra for all the postannealed samples. Only the nonpreannealed one shows a magnetic hyperfine field with an isomer shift equal to that of $\alpha$-Fe. The value of the magnetic hyperfine field is distributed over a wide range so that it can be assumed that a large part of the Fe ions also does not contribute to the full magnetic bulk moment. In contrast, no indication for metallic Fe exists in the spectra of the preannealed samples. They show similar hyperfine parameters dominated by a $Fe^{3+}$ doublet. Please note that after postannealing, $Fe^{2+}$ states are only present for the preannealed crystals but not for the nonpreannealed ones. The XAS measurements of the postannealed samples yield similar results (Fig. 3), i.e., ionic $2^+$ and $3^+$ valence states in all of the crystals with a contribution from metallic Fe solely in the nonpreannealed sample. Also for the $O_2$ preannealed sample we find a good coincidence between the Mößbauer and XAS data. That is, from the multiplet structure of the corresponding Fe $L_{2,3}$ XAS (third spectrum from the top in Fig. 3) one can conclude that $Fe^{3+}$ ions are dominating in this sample, whereas the presence of some $Fe^{2+}$ ions cannot be excluded. We find quite good agreement with the Fe $L_{2,3}$ XAS of $Fe_3O_4$ comprising 66.7% $Fe^{3+}$ and 33.3% of $Fe^{2+}$ ions. On the other hand, we find some differences in detail in the case of the vacuum preannealed crystal. The bulk sensitive CEMS suggests a very similar valence state than for the $O_2$ preannealed sample, dominated by $Fe^{3+}$ ions. The more surface sensitive XAS also suggests a mixed valence state, however, involving some more $Fe^{2+}$ than $Fe^{3+}$ states. The XAS of the $O_2$ preannealed sample is very similar to that of a $Sr_2FeMoO_6$ sample which has been found to have a mixed iron valence state involving around 65% $Fe^{2+}$ ions and 35% $Fe^{3+}$ ions.[16] This discrepancy could be explained by different spatial distributions of the charge states for the different preannealing conditions. From this analysis, we conclude that nanoparticle formation is suppressed by both preannealing methods. The mechanism of the suppression is not yet completely clear. Removal of defects acting as nucleation centers or introduction of defects immobilizing the Fe ions might be an explanation.

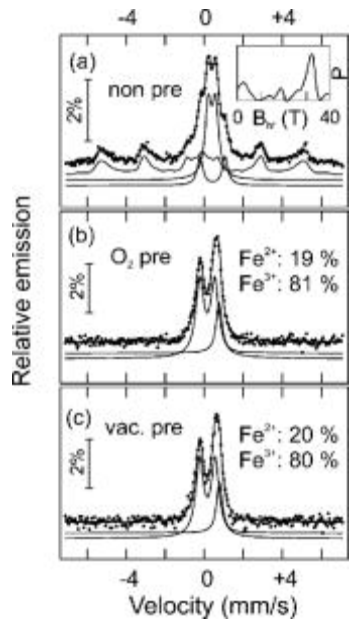

Fig. 2. CEM spectra and least squares fits with Lorentzian lines (see Ref. 15) of the Fe implanted and postannealed ZnO single crystals. (a) Nonpreannealed crystal with pronounced magnetic hyperfine field corresponding to α-Fe. The ratio between $Fe^0$ and $Fe^{3+}$ is 48%:52%, respectively. [(b) and (c)] CEM spectra of the preannealed samples. The ratio between the valence states is indicated.

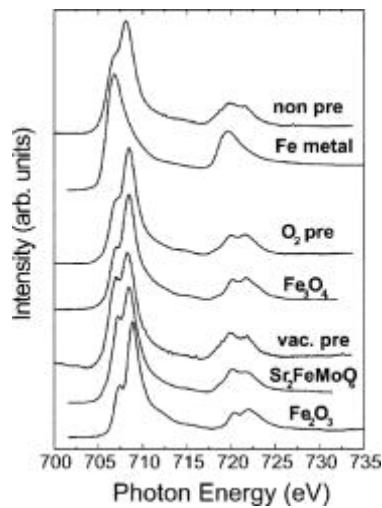

Fig. 3. Fe $L_{2,3}$ XAS of the nonpreannealed sample (top) and the two preannealed samples (third and fifth from the top) after implantation and postannealing. Several measurements on reference compounds, namely, Fe metal, $Fe_2O_3$, $Fe_3O_4$, and $Sr_2FeMoO_6$ are also shown for comparison (see Refs. 16,18). These spectra allow a qualitative determination of the Fe charge states in the ZnO samples. Please note that only the nonpreannealed sample shows pronounced contributions from metallic Fe.

The magnetic properties were analyzed by means of SQUID magnetometry. The hydrothermally grown virgin samples are purely diamagnetic with a susceptibility of $-2.6 \times 10^{-7}$ emu/g Oe. This value is consistent with the one observed by Quesada et al.,[17] i.e., $-1.62 \times 10^{-7}$ emu/g Oe. The difference might originate from the much different preparation method of the ZnO samples by this group. Pronounced ferromagnetic properties were only found for the nonpreannealed crystal after postannealing [Fig. 4(a)]. Magnetization reversal and zero field cooled (ZFC)/field cooled (FC) temperature dependence measurements recorded at 50 Oe show typical behavior of superparamagnetic nanoparticles with size distribution.[4] The nonpreannealed and the $O_2$ preannealed crystals do not show magnetic ordering for the as-implanted state (not shown). In contrast, after postannealing a weak separation between ZFC and FC curves up to 70 K can be observed for the $O_2$ preannealed crystal and up to a temperature above 250 K for the vacuum preannealed crystal [Figs. 4(b)4(c)]. Note that weak ferromagnetic properties occur already after implantation for the vacuum preannealed crystal [Fig. 4(d)]. The saturation magnetization extracted from hysteresis loops recorded at 5 K is below $0.025\mu_B$ per implanted Fe ion. As compared to $\alpha$-Fe the magnetic moment per implanted Fe ion is about 20 times smaller than the as-purchased crystal after postannealing. The shape of the ZFC-FC curve could be explained assuming regions with inhomogeneous Fe content as can be expected from the low temperature implantation. Postannealing, however, smoothes the ZFC-FC curve. The origin of the observed ferromagnetic properties is rather speculative at this point. First, due to the very low saturation magnetization achieved, we conclude that a large amount of defects created by high temperature annealing, probably oxygen vacancies, do not lead to pronounced ferromagnetic coupling of the implanted Fe ions. Second, it is rather likely that implantation or implantation plus mild postannealing creates such kind of defects, which lead to ferromagnetic properties of the Fe implanted ZnO. One possibility is the coupling of a small part of the Fe ions via Zn interstitials.

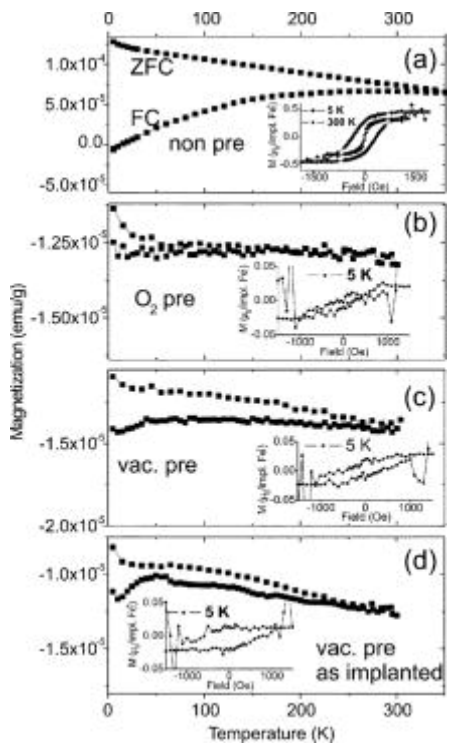

Fig. 4. ZFC-FC magnetization vs temperature measurements and magnetic hysteresis loops (insets) for all Fe implanted and postannealed ZnO single crystals [(a)–(c)]. The ZFC curves were obtained by cooling the sample from 300 down to 5 K in zero field and subsequently annealing it back to 300 K in 50 Oe field. The FC curves were obtained during subsequent cooling of the sample down to 5 K in a 50 Oe field. For the insets, the diamagnetic background was subtracted. (a) Nonpreannealed sample exhibiting $\alpha$-Fe nanoparticles [(b) and (c)] $O_2^-$ and vacuum preannealed crystals after postannealing. (d) As-implanted vacuum

preannealed crystal (for comparison). The latter three show a weak separation in the ZFC-FC and very low saturation moment in the hysteresis loops, as compared to (a). For (c) and (d), the thermomagnetic irreversibility temperature is above 250–300 K.

In summary, we demonstrated that preannealing of commercial ZnO(0001) single crystals in both flowing $O_2$ or vacuum suppresses metallic secondary phase formation after Fe implantation and mild postannealing in contrast to the nonpreannealed crystals. Weak ferromagnetic properties are induced in the vacuum preannealed crystals. These properties cannot be associated with ordinary superparamagnetic nanoparticles but could be due to indirect coupling mediated by point defects.